\documentclass[aps,pre,twocolumn,10pt]{revtex4-1}
\usepackage[utf8]{inputenc}
\usepackage[english]{babel}
\usepackage{amsmath}
\usepackage{amsfonts}
\usepackage{amssymb}
\usepackage[usenames,dvipsnames]{xcolor}
\usepackage{graphicx}
\usepackage{hyperref}
\usepackage{ulem}
\hypersetup{
  colorlinks   = true, 
  urlcolor     = magenta, 
  linkcolor    = red, 
  citecolor    = blue 
}

\newcommand{\bsl}{\backslash}
\newcommand{\mh}{\mathbf}
\newcommand{\mb}{\mathbb}
\newcommand{\mc}{\mathcal}

\newcommand{\tn}{\textnormal}
\newcommand{\bs}{\boldsymbol}
\newcommand{\bschi}{\boldsymbol{\chi}}

\begin{document}
\title{Anomalous dynamics and the choice of Poincar\'e recurrence-set}
\author{Matteo Sala$^{1}$}
\email{matteo.sala.teo@gmail.com}
\author{Roberto Artuso$^{2,3}$}
\email{roberto.artuso@uninsubria.it}
\author{Cesar Manchein$^{1}$}
\email{cesar.manchein@udesc.br}
\affiliation{$^{1}$Departamento de F\'\i sica, Universidade do Estado
  de Santa Catarina, 89219-710 Joinville, (Brazil)} 
\affiliation{$^{2}$Center for Nonlinear and Complex Systems and
  Dipartimento di Scienza ed Alta Tecnologia, Via Valleggio 11, 22100
  Como (Italy)} 
\affiliation{$^{3}$I.N.F.N., Sezione di Milano, Via Celoria 16, 20133
  Milano (Italy)} 
\begin{abstract}
 We investigate the dependence of Poincar\'e  recurrence-times statistics
 on the choice of recurrence-set, by sampling the dynamics of two- and
 four-dimensional Hamiltonian maps.
 We {derive a method} that allows us to visualize the direct relation
 between the {\it shape} of a recurrence-set and the {\it values} of its
 return probability distribution in arbitrary phase-space dimensions.
 Such procedure, which is shown to be quite effective in the
 detection of tiny regions of regular motion, allows to explain
 {\it why} similar recurrence-sets have very different distributions and
 {\it how} to modify them in order to {\it enhance} their return probabilities.
 {Applied on data}, this permits to understand the co-existence of extremely
 long, {\it transient} power-like decays whose anomalous exponent {\it depends}
 on the chosen recurrence-set. 
\end{abstract}

\maketitle

\section{\label{intro} Introduction}
One of the key issues in dynamical systems theory involves the
analysis of stochastic features manifested by chaotic deterministic
processes.
In this regard a fundamental property is represented by the decay of temporal correlation
functions: intuitively this is associated to loss of memory due to instabilities.
We remind that correlation decay is associated to mixing, a stronger property than
ergodicity in the hierarchy of chaotic indicators (see for instance \cite{aa89}).
When we focus on physical applications we are naturally
led to better characterize the way in which correlations
asymptotically vanish: while in the case of strongly chaotic systems
an exponential decay is expected (and proved, in a very same way of
space correlations for non critical lattice systems), the situation
is considerably more complicated when a chaotic sea coexists with regular regions.
In this case a typical chaotic trajectory gets
trapped for very long time segments close to regular structures,
significantly degrading the loss of memory in the dynamics.
Such a ``sticking'' phenomenon \cite{k83,bg90,szk93,sws93,mk00} quantitatively
leads to slow, polynomial decay of correlations (like in critical points of lattice models).
A sharp characterization of the power law exponent is crucial on many respects: 
on the one side a sufficiently small exponent leads to the violation of the central
limit theorem and might signal anomalous transport \cite{bg90,ab14} (since the diffusion
constant in Green Kubo formula diverges when correlations are not integrable); on the other
side, for important classes of Hamiltonian systems such exponents are conjectured to be universal \cite{ck08,s10}.
Direct calculation of correlation functions turns out to be a computationally
demanding task, especially when, like in aforementioned cases, we need to span several
decades in time in order to get clear indications of the appearance of power laws.\\

An alternative procedure, that introduces the main subject of
the present paper, involves a statistical analysis of the
{\it recurrence-times} of appropriate reference sets.
The mathematical result that justifies such an approach is the
Poincar\'e recurrence theorem \cite{p90}, which states that
{\it all} orbits of any volume-preserving transformation $\mh{f}$,
acting on a {\it bounded} phase-space $\Omega$ of any finite dimension,
eventually return arbitrarily near to their initial condition.
More precisely, in any neighbourhood $U_{\mh{x}}$ of any point $\mh{x}\in\Omega$
there is a point $\mh{y}$ that will come back to $\mh{f}^\tau(\mh{y}) \in U_{\mh{x}}$
in some unknown but {\it finite} time $\tau$.
Once we fix a reference set $B\subset\Omega$ we may thus investigate the
{\it probability density} $\rho_B(\tau)$ of first recurrence-times (RT)
through the fraction of initial conditions inside $B$ that return to $B$,
{\it for the first time}, exactly after $\tau$ iterations.\\

The relationship between such quantity and the
general problems mentioned earlier lies in the fact that the product
$\tau \cdot\rho_B(\tau)$ has the same asymptotic behaviour of
correlation functions: this was early recognized \cite{k83,cs84} in
the analysis of ergodic properties of area-preserving mappings, and
the correspondence has been mathematically proved \cite{y98,y99}.
The effectiveness of this technique is witnessed by the large range of
significant problems that have been addressed by RT statistics: from
universal properties of Hamiltonian systems with mixed phase-space
\cite{ck08,am09,ks13,s10,smba15,ak07}, to dynamical properties of
comets in the solar system \cite{sh10} and three-dimensional 
non-Hamiltonian models for fluid flows \cite{sbm15}, to statistical
properties of DNA molecules \cite{ms15}.
We point out that the most remarkable examples refer to polynomial
decay of RT statistics (namely systems with long range memory):
the transition from exponential to power-law decay of $\rho_B(\tau)$
has been extensively investigated for two-dimensional billiard tables \cite{acg96}.\\

\begin{figure*}[!t]
  \centering
  \includegraphics[scale=0.8]{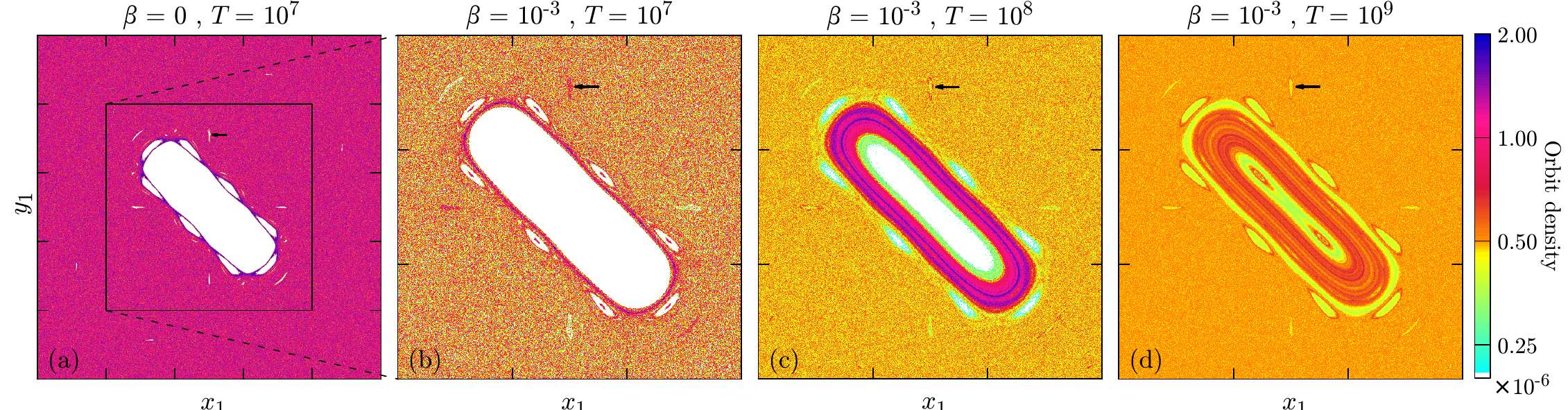}
  \caption{(Color online) Numerical orbit densities in the
    phase-space $(x_1,y_1)\in[0,1]^2$ of map (\ref{map2}) with
    $K_1=0.65\,,\,K_2=0.60$. Panel (a) is the uncoupled case $\beta=0$
    (single standard-map) after $T=10^7$ iterations; panels (b), (c)
    and (d) are magnifications of the square $[0.2,0.8]^2$ (depicted
    in panel (a)) for $\beta=10^{-3}$ respectively after $T=10^7$,
    $10^8$ and $10^9$ iterations. While the standard-map density
    exhibits {\it accumulations} (panel (a), purple/dark-gray
    cells) at the borders of stability-islands (white color
    indicates unvisited cells), the latter become accessible to the
    coupled map, although at extremely slow rates: at equal times
    (panel (b)), the system just started to penetrate areas {\it
      forbidden} to the standard-map. Only at much longer times
    $\sim10^9$ (panel (d)), phase-space exploration appears complete
    but anisotropies are still present. Such {\it filling} time-scale
    diverges in the limit of vanishing coupling $\beta\to0$, while
    there is a {\it threshold} $\beta_*<\beta$ above which
    map~(\ref{map2}) is completely ergodic \cite{kg87,kg88} (see
    Sec.~\ref{wct}).}  
  \label{fig2}
\end{figure*}

Though the asymptotic behaviour {of $\rho_B(\tau)$} at large RT's
is expected to be independent on the choice of recurrence-set
$B$, arbitrarily long transients may be present, endowed with
pre-asymptotic decay that may indeed depend upon the choice of $B$.
These features are of particular relevance in the investigation of
Hamiltonian systems with mixed phase-space, where, after a {\it transient},
the decay of {\it integrated} RT probabilities $\tn{P}_B(\tau)=\sum_{t\geq
  \tau}\rho_B(t)\sim\tau^{-\gamma}$ turns algebraic. While the asymptotic 
exponent $\gamma$ is assumed to be independent on $B$ (and, for 2D
Hamiltonian maps, there are arguments supporting its universality
$\gamma_{2\tn{D}}\simeq 1.6$ \cite{mo85,ck08,am09,ks13}), no general
statements are known {about} the transient behaviour {and its length} (precise
results on the full probability distribution are only available in
the limit of increasingly small recurrence-sets $B$, see
\cite{hsv99} and references therein).
When we increase the dimensionality $N>2$ there are a few numerical
investigations (for small and moderate numbers of {\it homogeneously} 
coupled 2D maps) that again suggest a (universal) asymptotic exponent
$\gamma_{N\tn{D}} \simeq 1.2$ \cite{s10,smba15,ms15}, but it is still
under debate whether such value persists in the weak-coupling regime.
In particular, it remains unclear how to estimate the {\it time-scale}
over which a higher-dimensional system reaches its asymptotic regime
under the process of {\it weak} Arnol'd diffusion \cite{a64,ll92}.
In this work we address such problem by linking the {\it diffusive strength}
along a specific coupling-{transition} across its {\it ergodic threshold} (see \cite{kg87,kg88}
and references therein) to the {\it dependence} of RT's on the choice of recurrence-set $B$.\\

The paper is organized as follows: while in Sec.~\ref{mo} we introduce
the four-dimensional model under study, in Sec.~\ref{dc} we describe
how recurrence data is generated and collected by our numerical
experiments, providing the due mathematical background through appendices.
On such grounds, in Sec.~\ref{res} we present our numerical results:
by the developed methods we have access to an efficient {\it detector}
for all phase-space structures that are responsible for {\it long}
recurrences (down to very small scales, see Sec.~\ref{sid}) which, in
turn, suggests a recipe to systematically {\it enhance} RT probabilities
by changing a specific {\it universal} property of the chosen set (namely,
the {\it size} of its {\it departure-set}, see Sec.~\ref{ai}).
By performing our novel RT analysis to several recurrence-sets $B$ in
comparing different parameters for system~\ref{mo}, in Sec.~\ref{wct}
we confirm the existence of {extremely} long sub-diffusive transients
with pre-asymptotic exponent $\gamma_B$ {\it dependent} on set $B$.
After exploiting the aforementioned detector to also {\it identify}
the phase-space structures responsible for such a non-trivial dependence,
in Sec.~\ref{conc} we summarize our results indicating some of their possible implications.

\section{Model}
\label{mo}
\begin{figure}[!b]
  \centering
  \includegraphics[scale=0.5]{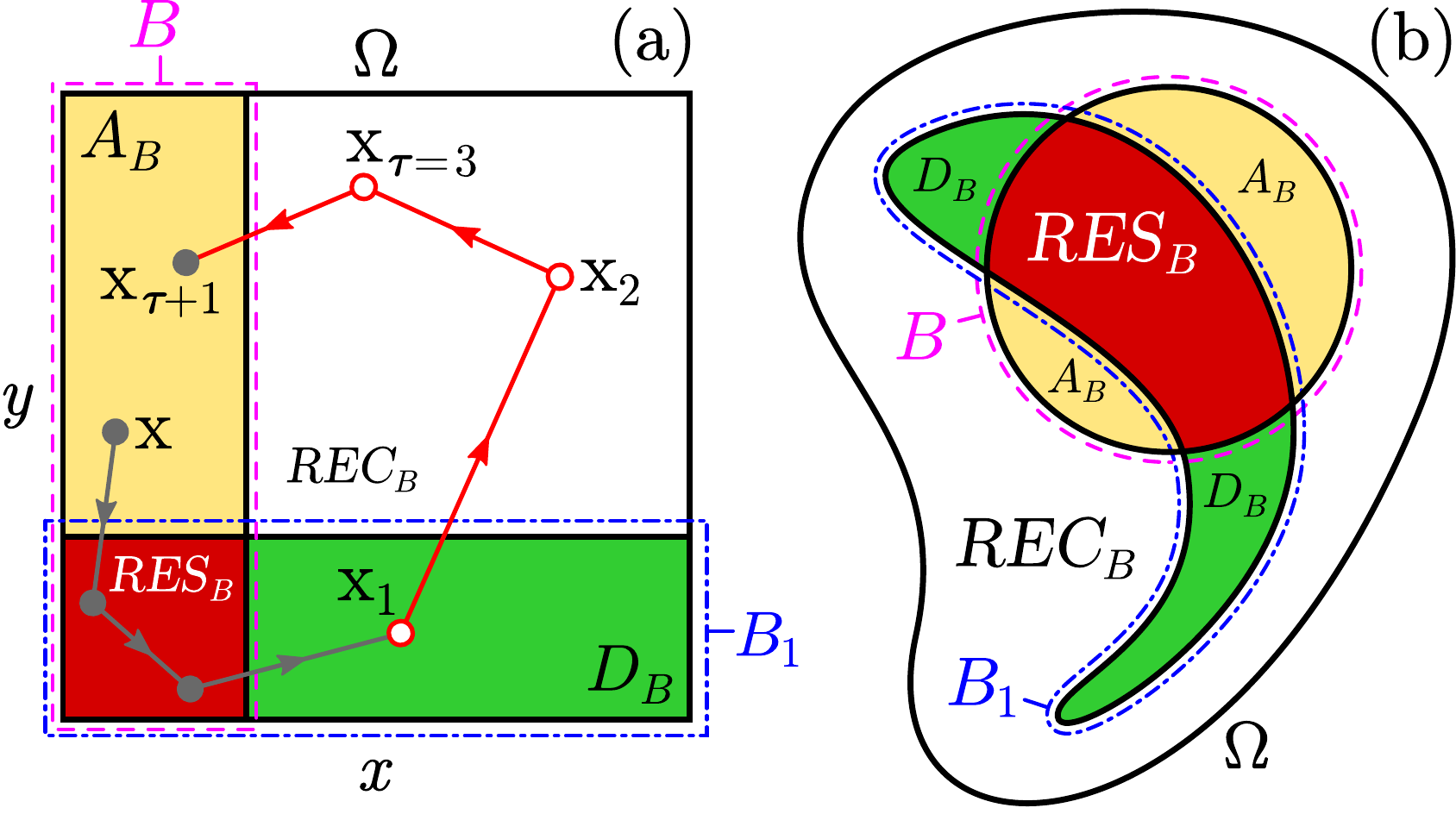}
  \caption{(Color online) Panel (a): scheme of a single Poincar\'e
    cycle as described in Sec.~\ref{dc} for one uncoupled ($\beta=0$)
    standard-map in (\ref{map2}) and recurrence-set $B$ (bordered by
    pink/gray dashed line) in the phase-space $\Omega=[0,1]^2$ (largest
    square), with departure-set $D_B=B_1\bsl B$ (green/gray zone) and
    arrival-set $A_B=B\bsl B_1$ (yellow/light-gray zone);
    recurrence/residence cycles take place only in the respective sets
    $REC_B / RES_B$. In these coordinates, first iterated-set 
    $B_1=\mh{f}(B)$ (bordered by blue/black dash-dotted line)
    has the same 
    shape of $B$ {\it reflected} about the bisector. Panel (b):
    picture of the four sets $D_B$, $REC_B$, $A_B$ and $RES_B$ as
    described in Appendix~\ref{DA} for a dynamics $B\to B_1=\mh{f}(B)$
    induced by evolving a generic flow $\mh{f}=\mc{F}^{\Delta t}$
    along a fixed time-step $\Delta t$; standard-map in panel (a)
    belongs to the special class of {\it time-periodic} Hamiltonian
    flows whose integration after one period $\Delta t$ admits an {\it
      explicit} expression, as the one used in equation~(\ref{map2}). }
  \label{fig1}
\end{figure}

We consider as benchmark system a pair of coupled 2D
Chirikov-Taylor standard-maps \cite{ch79}, defined on
the unit torus $(x_j,y_j)\in[0,1]^2=\mb{T}^2$ for
$j=1,2$ and evolving by:
\begin{align}
  \left\{
    \begin{array}{l}
      \bar{x}_j\ =\ 2\,x_j-y_j+f_j(x_1,x_2)\quad\tn{mod}\ 1\ ,\\ 
      \bar{y}_j\ =\ x_j\ ,
    \end{array}
  \right.
  \label{map2}
\end{align}
with both local ($K_j$) and coupling ($\beta$) {\it conservative}
non-linear forces which, setting $\delta x=x_1-x_2$, are given by:
\begin{align}
  f_j(x_1,x_2)=K_j\sin(2\pi\,x_j)+\beta\,(-1)^j\,\sin(2\pi\,\delta x)\ .
\end{align}
Map~(\ref{map2}) is four-dimensional {\it symplectic} (phase-space is $\Omega=[0,1]^4=\mb{T}^4$),
with single 2D standard-maps written in McMillan form \cite{m67} to fully exploit
their symmetry structure (see the orbit densities in Fig.~\ref{fig2}). Indeed, in such
coordinate system, each 2D standard-map $(x,y)$ in~(\ref{map2}) at $\beta=0$ maps vertical lines
$x=x_0$ into their symmetric horizontal lines $y=x_0$ {\it independently}
from the specific local force $K\sin(2\pi x)$, which only sets a {\it shift} along the reflected
line itself.
Interestingly, all vertical lines inside the rectangular set $B$
sketched in Fig.~\ref{fig1} (bordered by pink/gray dashed line)
are thus reflected, but also non-linearly shifted along themselves and
wrapped over the torus; as a consequence, set $B$ is transformed into
$B_1=\mh{f}(B)$ (bordered by blue/black dash-dotted line) whose
{\it shape} coincides with $B$ reflected about the bisector.\\ 

Although in different coordinates, the same system is studied in \cite{kg88}
for several numbers of {\it identical} coupled maps $K_j=K\,,\,\forall\,j$,
confirming the existence of a coupling {\it threshold} $\beta>\beta_*$ above
which dynamics appears ergodic.
A similar interaction force $\pm\beta\sin(2\pi(\,x_1+x_2))$ is also considered in
\cite{rsbk14}, finding a non-trivial structure of {\it resonances}
between the two 2D standard-maps.
Local forces $K\sin(2\pi x)$ ensure that system~(\ref{map2})
is chaotic also when it is uncoupled, at $\beta=0$.
To get a generic case, we fix the parameters $K_1=0.65$ and $K_2=0.60$ \cite{Note1},
both corresponding to well-developed chaotic regimes with strongly {\it mixed}
phase-spaces, due to the co-existence of a main hyperbolic fixed-point at 
$(x,y)=(0,0)$ with a central system at $(x,y)=(0.5,0.5)$.
While for $K_2=0.60$ the latter is a stable fixed-point with a typical surrounding hierarchy
of un/stable tori, for $K_1=0.65$ it is made by a {\it weakly} hyperbolic fixed-point with
{\it eight-shaped} homoclinic intersections encircling a {\it stable} orbit of period 2.\\
Remarkably, in the uncoupled regime (Fig.~\ref{fig2}~(a), $\beta=0$) the central system is
{\it completely} isolated from the main chaotic sea by a thick, quasi-periodic annular barrier. 
Once the  coupling is on (Fig.~\ref{fig2}~(b)-(d) for $\beta=10^{-3}$), the orbits are allowed
to penetrate the barrier and access the central system, although at gradually slower rates as
$\beta\to0$, {\it de facto} inducing the process of {\it weak} Arnol'd diffusion.
Operatively, by the coupling parameter-range $10^{-6}\leq\beta\leq10^{-3}$
we observe weakly {\it correlated} kicks, given by $\pm\beta\,\sin(2\pi\,\delta x)$,
driving each of the two maps in and out what become {\it broken} tori.
The latter appear to be alternatively selected/avoided by the orbits in a very complicated way,
which sensibly depends on the initial condition: an example is shown in Fig.~\ref{fig2}~(b)-(d)  
by a single orbit that accumulates more over {\it just} one of two {\it symmetric} period-3
island-chains with very thin shape (the arrow points to one of the three selected islands).
By changing initial condition (not shown) it is possible to revert such {\it sticky}
behaviour to the other period-3 island-chain, or even to {\it avoid} it, highlighting
the existence of markedly {\it separate} paths by which orbits can penetrate and
diffuse inside the broken central system.

\section{Data collection}
\label{dc}
The standard procedure to collect a single event $\tau$
in {the} RT statistics involves what { is called} a {\it Poincar\'e cycle} \cite{z95}.
In spite of being a very old recipe, this deserves to be examined with care,
more than for clarity, for the useful information it encodes. Informally, indeed,
one can describe the procedure as: measure the lapses of time spent outside set $B$
by an orbit started in $B$.
Here we focus on the single recurrence cycle and, by referring to Fig.~\ref{fig1},
picture it as follows:
an orbit starts from point $\mh{x}$ in set $B$ (bordered by
pink/gray dashed line) and evolves as long as it remains {{\it
    inside}} $B$ (solid bullets). 
The {\it residence} period inside set $RES_B$ (dark-red/dark-gray zone) lasts
until landing in what we call the {\it departure-set} $D_B$
(green/gray zone), made by all points $\mh{x}_1$ that escape
from $B$ in {\it one} iteration into $B_1=\mh{f}(B)$ 
(bordered in blue/black dash-dotted line).
Notice that this new set is constructed by $D_B=B_1\bsl B$, {\it i.e.} by
subtracting from set $B_1$ all points that are still in $B$.
From such first point $\mh{x}_1\in D_B$ lying outside $B$, we start a clock
$t_B(\mh{x}_1)=1$ and employ it to label all successive iterates
$\mh{x}_t\notin B$ that stay outside $B$, in $REC_B$ (white zone) by
setting $t_B(\mh{x}_t)=t$ (red/empty circles). A cycle ends once
the orbit returns to $B$ after a number $t=\tau$ of clock ticks, 
which marks its Poincar\'e RT (length).
By repeating the substitution $\mh{x}\leftarrow\mh{x}_{\tau+1}\in B$ a number $E$ of times,
we thus collect RT statistics over an ensemble made of $E$ Poincar\'e cycles
(events) sampled from a single trajectory started in $B$ (see Appendix~\ref{hs}).\\

All points $\mh{x}_{\tau+1}$ hitting $B$ first belong to what we call
the {\it arrival-set} $A_B$ (yellow/light-gray zone in
Fig.~\ref{fig1}) which is the {\it only} set accessible from outside
$B$ in one step. 
Notably, the arrival-set has a general expression $A_B=B\bsl B_1$ which is similar
to the departure-set $D_B=B_1\bsl B$ and shares with it the same volume $\mu(A_B)=\mu(D_B)$.
By fixing set $B$, any bounded phase-space $\Omega$ is then
partitioned into four sub-sets $D_B$, $REC_B$, $A_B$ and $RES_B$
respectively associated to the {\it universal} processes of:
departure, recurrence, arrival and {\it residence} in $B$ (in
Fig.~\ref{fig1}, green/gray, white, yellow/light-gray
and dark-red/dark-gray zones), whose shape is induced by both
the dynamics and the recurrence-set $B$. 
This suggests that the departure/arrival-sets $D_B$ and $A_B$ are the {\it unique}
channels between escape and return cycles, which practically means that the departing
volume $\mu(D_B)=\mu(A_B)$ measures the {\it rate} per single iteration at which a
volume-preserving dynamics extracts/re-injects orbits from a set $B$ (see Appendix~\ref{DA}).\\

Such mechanism is better understood by inspection of the employed clock
$t_B(\mh{x})$, which we name {\it Poincar\'e-time} (PT): this is the
time $t_B(\mh{x})$, since {\it last} exit, spent by a Poincar\'e cycle to reach point
$\mh{x}\in \Omega\bsl B$ lying {\it outside} $B$.
In case $\mh{f}$ is a mapping, PT is an integer-valued {\it function} which ranges from $1$ to $\tau$
when observed over a cycle of length (RT) equal $\tau$.
Consequently, once PT is evaluated over ensembles of cycles ({\it i.e.} volumes outside $B$),
the PT level-sets $S^\tau_B$, made of points $\mh{x}$ with {\it constant} PT value:
\begin{align}
S^\tau_B\ :=\ \{\ \mh{x}\in \Omega\bsl B\quad\tn{s.t.}\quad t_B(\mh{x})=\tau\ \}\ ,
\label{PTlev}
\end{align}
reveal the shapes of all the higher iterates $B_\tau=\mh{f}^\tau(B)$ of set $B$.
Indeed, in Appendix~\ref{ptf} we show how the whole set-family $\{S_B^\tau\}_{\tau=1,2,..}$
can be generated by the universal recursive relation $S^{\tau+1}_B=\mh{f}(S^\tau_B)\bsl B$ with
departure-set $S^1_B=B_1\bsl B= D_B$ as initial condition at PT $\tau=1$.
In conjunction with Appendix~\ref{cp}, we also prove that all the volumes $\mu(S^\tau_B)$
of the PT level-sets family can be used to express the full RT probability
$\tn{P}_B(\tau)=\tn{Prob}(\tn{RT}\geq\tau)$:
\begin{align}
  \tn{P}_B(\tau)\ =\ \mu\left(S^\tau_B\right)\ /\
  \mu\left(D_B\right)\ .
  \label{probok1}
\end{align}
by {\it normalizing} them exactly by the departing volume $\mu(D_B)$, as suggested by our
preliminary considerations.
Novel expression~(\ref{probok1}) highlights a fundamental fact: RT probabilities
decay {\it exactly} as the PT level-sets volumes $\mu(S_B^\tau)\to 0$, {\it weighted}
by the departing volume $\mu(D_B)$.
While the latter acts as a pre-factor to the
asymptotic decay (which is assumed to be independent on set $B$), formula (\ref{probok1})
already shows that {\it transient} RT statistics is entirely implied by the {\it relative} size
of PT level-sets $S_B^\tau$ for $\tau=2,3,4,..$ with respect to the first departure-set $S_B^1\equiv D_B$.
Such universal property is the key ingredient upon which we base our analysis of
the dependence of RT statistics on the choice of recurrence-set.

\section{Results}
\label{res}
The considerations from previous section~\ref{dc} (see also the Appendices
for more detailed discussions) find their application in two fundamental aspects
of the same problem: RT probabilities $\tn{P}_B(\tau)$ depend on both the
{\it intrinsic} structure of phase-space (induced by dynamics) and the
{\it external} choice of recurrence-set (chosen by observers).
Poincar\'e-time functions (PT) provide a rigorous
method to {\it chart} and identify all the {\it localized} sources of long
RT's, for all the conservative systems in higher-dimensions.
By employing PT as a {\it detector} for extremely small phase-space structures 
(see Sec.~\ref{sid}), we also find a simple recipe which allows to {\it shift}
in $\tau$ the RT probability $\tn{P}_B(\tau)$ by appropriately changing the {\it shape}
of recurrence-set $B$ (see section~\ref{ai}). Finally, by combining such findings,
we are able to assess a {\it lower-bound} for the transient time-scale of RT statistics (see section~\ref{wct})
for the four-dimensional system presented in Eq.~(\ref{map2}).

\begin{figure}[!b]
  \centering
  \includegraphics[scale=0.73]{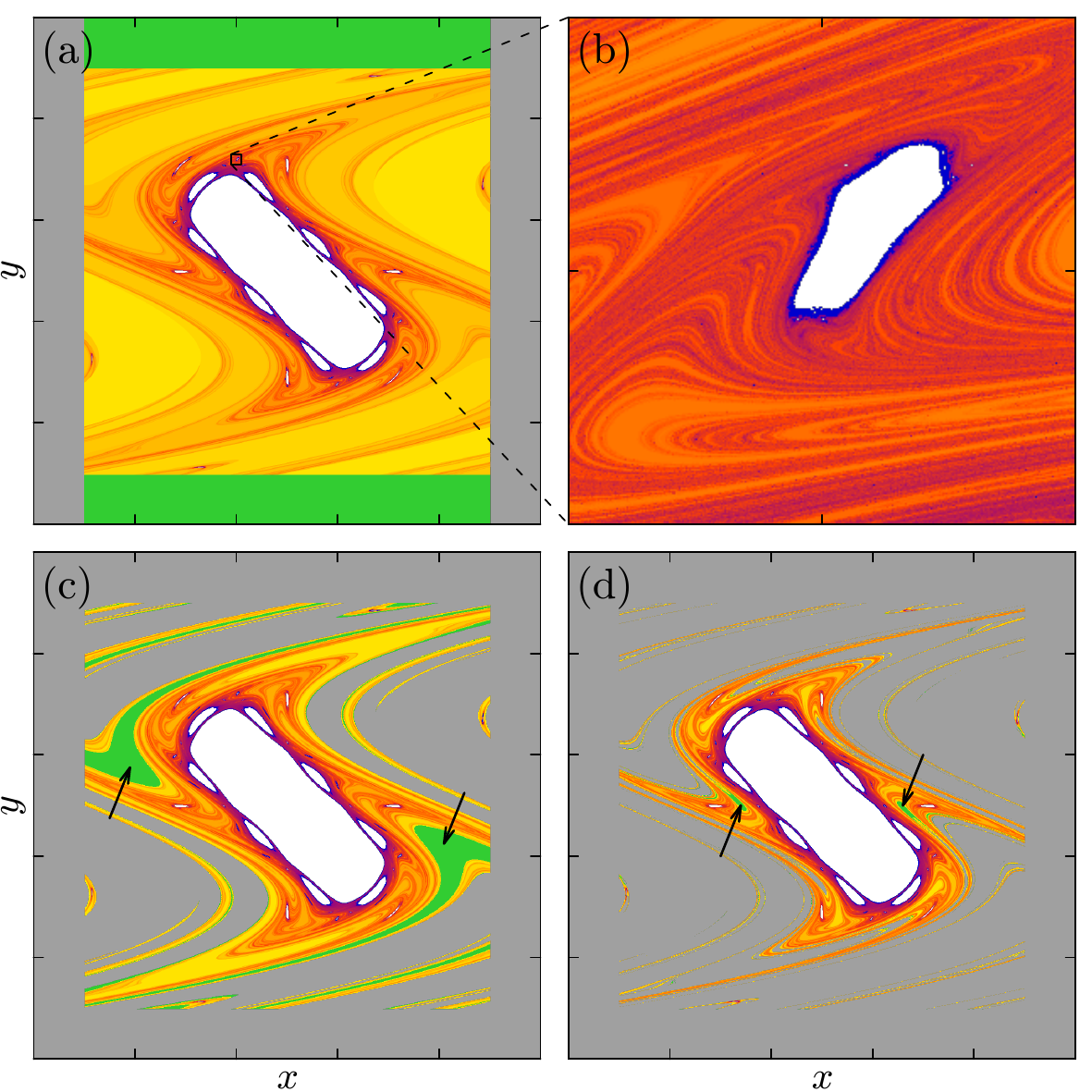}
  \caption{(Color online) Panel (a): Poincar\'e-time (PT) function for
    the same uncoupled standard-map as in Fig.~\ref{fig2}~(a) with
    recurrence-set $B_{[0.1]}=\{|x_1|<0.1\}$ (gray rectangles),
    departure-set $D_B$ (green/dark-gray rectangles) and PT
    level-sets $S^\tau_B$ as in Eq.~(\ref{solset}) with colors encoded
    by the $\tau$ axis in Fig.~\ref{fig4}; white color indicates
    cells {\it unvisited} by Poincar\'e cycles. Panel (b):
    magnification of the square $[0.39,0.41]\times[0.71,0.73]$ from
    panel (a), detecting islands structures of extremely small area
    $\sim10^{-8}$. Panels (c) and (d): same PT analysis as panel (a)
    for two {\it lag-sets} $I^{\delta\tau}_{B}$ as in Eq.~(\ref{iset})
    (gray cells) respectively for $\delta\tau=5$ and $10$.
    The lag-sets departing volume $\mu(D_{I^{\delta\tau}_{B}})$
    (green/dark-gray cells indicated by arrows) {\it shrinks}
    as we rise $\delta\tau$.} 
  \label{fig3}
\end{figure}

\subsection{Phase-space structures detection}
\label{sid}
\begin{figure}[!t]
  \centering
  \includegraphics[scale=0.85]{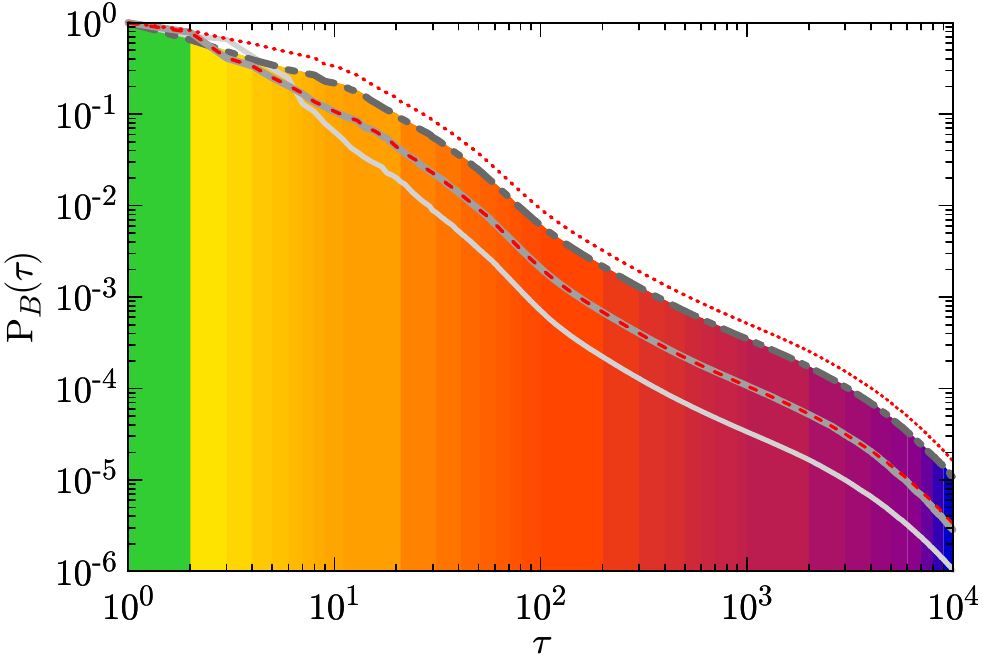}
  \caption{(Color online) RT probability (light-gray continuous
    curve) for the recurrence-set $B$ in Fig.~\ref{fig3}, panel~(a)
    and its {\it lag-sets} $I^{\delta\tau}_{B}$ with lags
    $\delta\tau=5$ (dashed red/black curve, set in
    Fig.~\ref{fig3}~(c)) and $\delta\tau=10$ (dotted red/black
    curve, set in Fig.~\ref{fig3}~(d)) as in Eq.~(\ref{probokI}). The
    predicted probabilities for $I^{\delta\tau}(B)$ are matched in the
    case $\delta\tau=5$ (continuous gray curve for the set
    $I^5(B_{[0.1]})$ in Fig.~\ref{fig3}~(c)) while they are
    underestimated in the case $\delta\tau=10$ (dash-dotted
    dark-gray curve for the set $I^{10}(B_{[0.1]})$ in
    Fig.~\ref{fig3}~(d)). Despite this, both cases produce RT
    probability {\it enhancement}.}  
  \label{fig4}
\end{figure}

In figure~\ref{fig3}, we start by applying expression~(\ref{probok1})
for RT statistics to the {single} standard-map $K=0.65$, which corresponds
to the uncoupled regime of four-dimensional map~(\ref{map2}) at $\beta=0$.
In panel (a), the unit-square phase-space $(x,y)\in[0,1]^2$ contains the recurrence-set
$B_{[0.1]}=\{|x|<0.1\}$ (gray rectangles) and its departure-set $D_B=B_1\bsl B$
(green/dark-gray rectangles) as discussed in Sec.~\ref{dc} and
defined in Appendix~\ref{DA}. 
The PT level-sets $S^\tau_{[0.1]}\equiv S^\tau_{B_{[0.1]}}$ for higher values of $\tau$
are approximated \cite{Note0} by the $\mc{G}$-averaged PT function $\langle t_B\rangle_\mc{G}$
computed over a square-grid $\mc{G}$ with $10^3\times10^3$ cells by the recipe
described in Appendix~\ref{ptf}.
As explained there, the PT probability-density {\it coincides}
with the RT probability, so that each color in Fig.~\ref{fig3} encodes a PT value which
corresponds to the color-scale along the $\tau$ axis in Fig.~\ref{fig4} while the value of
$\tn{P}_B(\tau)$ can be also interpreted as the {\it fraction} of cells with such color
(PT value) among grid $\mc{G}$.\\

At once, this allows to chart and visualize the natural {partition} of
phase-space induced by the association of each PT level-set (color) to
a specific time-scale in the RT statistics: range $\tau<10^2$ (colors
from yellow/light-gray to dark-orange/gray) is associated to the
chaotic-sea generated by the main unstable fixed-point at
$(x,y)=(0,0)$ while the ranges $10^2\leq\tau<10^3$ (from
dark-orange/gray to dark-magenta/dark-gray) and $10^3\leq\tau<10^4$ (from
dark-magenta/dark-gray to blue) appear to be generated by
{sub-diffusion} respectively through a {\it symmetric pair} of very
{thin} period-3 island-chains (one of which is indicated by arrows in
Fig.~\ref{fig2}) and the barrier made by a single period-8
island-chain. The highest recorded values of $\tau\sim10^4$ (bins in
blue color in Fig.~\ref{fig4}) are associated to the {\it
  sticky motion} around the main stable-island and are located in the
distribution tail, exactly where RT probabilities
$\tn{P}_B(\tau)\sim10^{-6}$ approach the cell-area of grid $\mc{G}$;
the latter is thus {\it insufficient} to resolve the phase-space
details of higher PT level-sets.\\

Indeed, by fixing a time-scale $\tau>\tau^*$ one gets a {\it unique} probability-threshold
$\tn{P}_B(\tau)<\tn{P}_B^*$ which, by Eq.~(\ref{probok1}), automatically implies an
{\it upper-bound} for each level-set volume $\mu(S^\tau_B)<\mu(S^1_B)\times \tn{P}_B^*$.
This makes the cell-averaged PT function $\langle t_B\rangle_\mc{G}$ a fast
{\it presence detector} for phase-space regions responsible for very long RT's
(such as micro-islands in strongly chaotic backgrounds) based on the {\it time-scale}
$\tau$ at which each of such structures contributes to the RT statistics.
A direct example of these considerations is shown in Fig.~\ref{fig3}~(b) by a magnification of panel (a)
by a grid $\mc{G}'$ with $300\times300$ cells over the square $(x,y)\in[0.39,0.41]\times[0.71,0.73]$:
the higher iterates $B_t$ of set $B$ naturally tend to align to the unstable manifolds as $t\to\infty$
\cite{Note2}, providing their structure in the neighbourhood of resonances {\it while} localizing
smaller islands, possibly for further magnifications.
Notice that this method outperforms existing detectors both in the
{observable} scales {\it and} computational {cost} \cite{tl07,Note3}.

\subsection{RT Probability enhancement}
\label{ai}
Among their possible applications in data analysis \cite{Note4}, PT
functions allow to perform a direct {comparison} between the different {\it sizes}
of localized RT sources.
In particular, from Sec.~\ref{dc} and Appendix~\ref{ptf} we know that the departing volume
$\mu(D_B)$ fixes the {\it maximal} size $\mu(D_B)\geq\mu(S^\tau_B)$ of
all the PT level-sets $S^\tau_B$ generating the RT statistics $\tn{P}_B(\tau)$,
whose decay is completely  determined by the {\it relative} size between
sets $S^{\tau}_B$ and $S^1_B\equiv D_B$.
If the latter is much {\it larger} than the formers, at short recurrences
$\tau_*$ the RT probability $\tn{P}_B(\tau<\tau_*)$ {\it falls} abruptly.
Conversely, by opting for recurrence-sets $B$ whose departing volume $\mu(D_B)\ll1$
is sufficiently small to have the first $\tau_*$ level-sets volumes
$\mu(S^{\tau\,<\,\tau_*}_B)\simeq\mu(S^1_B)$ {\it comparable},
we get an automatic {\it raise} of the first $\tau_*$ RT probabilities.
This is a consequence of the fact that small departing volumes $\mu(D_B)=\mu(B_1\bsl B)\simeq 0$
imply to have $B\simeq \mh{f}(B)$, that is, recurrence-sets that are
{\it almost-invariant}. In fact, the limit case of {\it invariant-sets}
$B=B_1$ saturates to 1 all RT probabilities: $\tn{P}_B(\tau)=1$ for all $\tau$ and $B$.\\

\begin{figure*}[!t]
  \centering
  \includegraphics[scale=0.9]{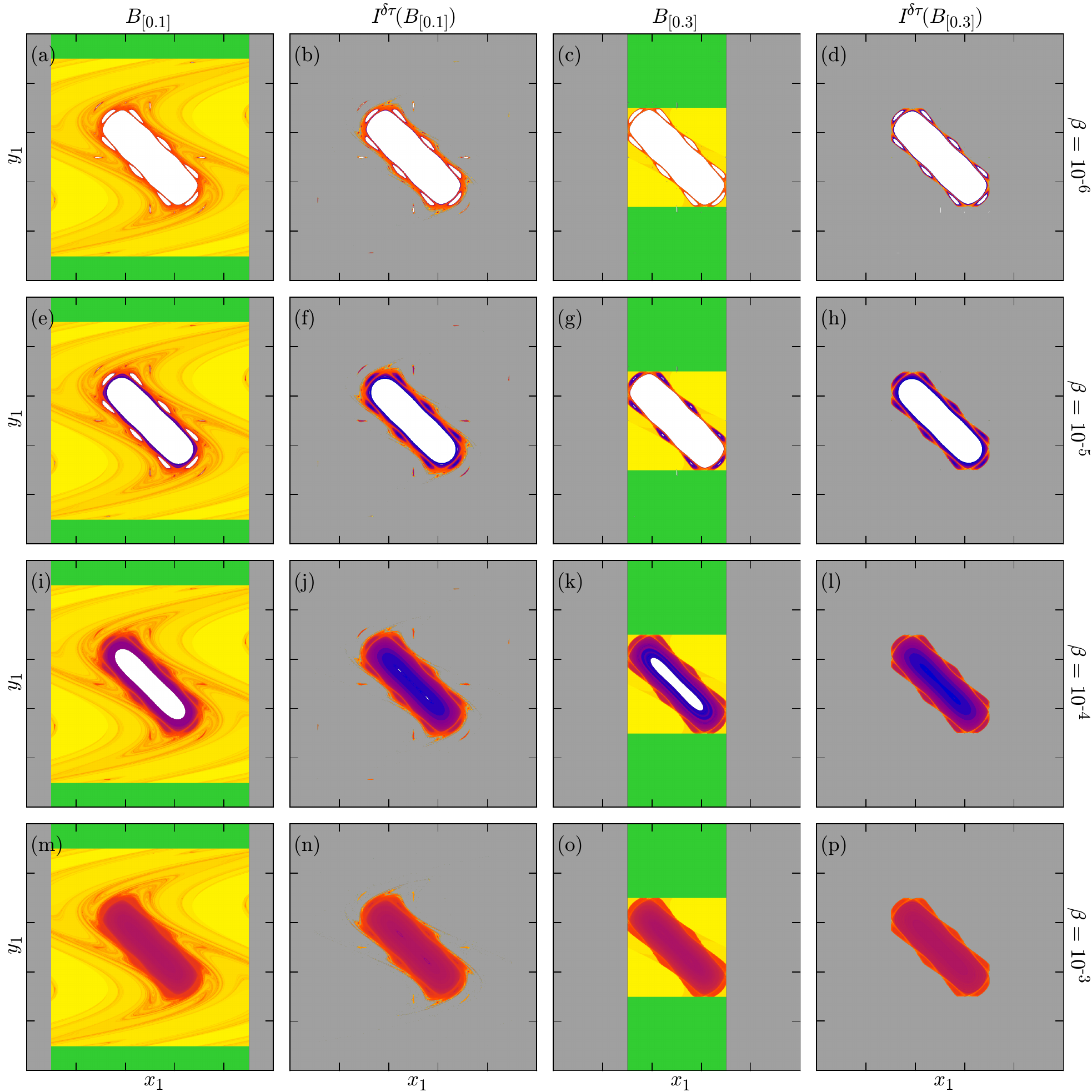}
  	\caption{(Color online) Same Poincar\'e time (PT) function analysis
  	as in Fig.~\ref{fig3} in phase-space $(x_1,y_1)=[0,1]^2$ for the
  	first of the two maps in Eq.~(\ref{map2}) at four coupling regimes:
    $\beta=10^{-6}$, $10^{-5}$, $10^{-4}$ and $10^{-3}$ (respectively
    from first to fourth row) for the recurrence-sets $B_{[0.1]}=\{|x_1|<0.1\}$
    and $B_{[0.3]}$ and their lag-sets $I^{\delta\tau}(B_{[0.1]})$ and
    $I^{\delta\tau}(B_{[0.3]})$ (gray zones, respectively from first
    to fourth column) with lag-time $\delta\tau=200$ in Eq.~(\ref{iset}).
    Colors are encoded by the $\tau$ axis in Fig.~\ref{fig5}, as
    $\mu(S^\tau_B)\propto\tn{P}_B(\tau)$ by Eq.~(\ref{probok1}), with
    green/dark-gray color for $D_B$ (for lag-sets $I^{\delta\tau}$, the
    latter is made of a few cells) and white color for unvisited
    cells.} 
  \label{fig6}
\end{figure*}

Inspired by such ideas and to avoid expensive procedures to find and test
new sets which are {\it also} almost-invariant, we define a family of sets
as {\it automatic} extensions of any, possibly already tested, set $B$.
These sets are made by {\it joining} to set $B$ all its iterates
$B_{t=1,2,..,\delta\tau}$ until some {\it lag-time} $\delta\tau$.
Accordingly, we name these the $\delta\tau$-th {\it lag-sets} $I^{\delta\tau}(B)$
acting over an arbitrary set $B$: 
\begin{align}
  I^{\delta\tau}(B)\ =\ \bigcup_{t=0}^{\tau}\,B_t\ . 
 \label{iset}
\end{align}
This construction is {strongly} motivated by the fact that each new PT level-set
$S^\tau_{I^{\delta\tau}(B)}=S^{\tau+\delta\tau}_B$ for set $I^{\delta\tau}(B)$
is {\it exactly} the old $(\tau+\delta\tau)$-th PT level-set for set $B$, which 
implies a straightforward relation between the {\it old} $\tn{P}_B(\tau)$ and the
{\it new} $\tn{P}_{I^{\delta\tau}(B)}(\tau)$ RT probability: 
\begin{align}
  \begin{array}{ll}
    \tn{P}_{I^{\delta\tau}(B)}(\tau)\ &=\
    \mu\left(\ S^{\tau+\delta\tau}_B\ \right)\ /\
    \mu(\ S^{1+\delta\tau}_B\ )\\\,
    \\
    &=\ \tn{P}_B(\tau+\delta\tau)\ /\ \tn{P}_B(1+\delta\tau)\ .
  \end{array}
  \label{probokI}
\end{align}
This is akin to say that the {\it lagging} of a recurrence-set $B\to I^{\delta\tau}(B)$
{\it shifts} by $\delta\tau$ its RT statistics while {\it scaling} it by $\tn{P}_B(1+\delta\tau)$.
In Fig.~\ref{fig3} we apply such recipe for two values of lag-time $\delta\tau=5$
and $10$ (respectively panel (c) and (d), white cells), by {\it ``turning white''}
all cells in panel (a) that, for rectangular set $B_{[0.1]}$, have $\mc{G}$-averaged PT function
$\langle t_{B_{[0.1]}}\rangle_\mc{G}\leq\delta\tau$ {\it below} the respective lag-time $\delta\tau$.
While any lag-set $I^{\delta\tau}(B)$ is always {\it bigger} than $B$ (otherwise
no point would ever exit from $B$), we confirm that the corresponding
departing volume is {\it smaller} and it {\it shrinks} for longer lags (see arrows in
Fig.~\ref{fig3}~(c) and (d)), since the PT level-set volumes $\mu(S^{\delta\tau+1}_B)\to0$
are {\it monotonic decreasing} in the lag-time $\delta\tau\to\infty$. 
Alongside, in Fig.~\ref{fig3}~(d) we also notice very {\it
  filamentary} structures in the shape of set $I^{10}({B_{[0.1]}})$,
with details having size {\it below} the cell-area scale. 
Unsurprisingly, the corresponding RT probabilities shown in
Fig.~\ref{fig4} reveal that only the first case matches the theory
($\delta\tau=5$, continuous gray curve \textit{vs}
Eq.~(\ref{probokI}), dashed red/dark-gray curve) while the
second case ($\delta\tau=10$, dash-dotted gray curve) {\it
  underestimates} prediction (\ref{probokI}) (dotted
red/dark-gray curve) because of the {\it finite-size} of our
lag-sets approximation. 
By the latter, we are thus able to {\it control} the departing volume
$\mu(D_{I^{\delta\tau'}_B})\leq\mu(D_{I^{\delta\tau}_B})$ and progressively enhance
the RT probabilities by employing larger lag-times $\delta\tau'>\delta\tau$,
but there is a maximal {\it effective} lag $\delta\tau_*$ which limits the
observations that are made at {\it fixed} resolution of square-grid $\mc{G}$.
Remarkably, indeed, the formal {\it infinite-lag} limit
$\delta\tau\to\infty$ for Eq.~(\ref{probokI}), saturates to 1 the RT
probability $\tn{P}_{I^{\delta\tau}(B)}(\tau)\to1$ for all $\tau$ and
sets $B$, so that $I^{\infty}_B$ is the invariant-set covering the
whole volume that is {\it accessible} from the chosen set $B$.

\subsection{Weak-chaos transition}
\label{wct}
Once testing the lag-set recipe described in Sec.~\ref{ai} on the
four-dimensional map (\ref{map2}) for $\beta>0$ we must take into
account that, in addition to the coarse-grain induced by square-grid
$\mc{G}$, the lag-sets are approximated through the relation:
$I^{\delta\tau}(B)\simeq\{\langle t_B\rangle_\mc{G}\leq\delta\tau\}$,
with the $\mc{G}$-averaged time-of-flight $\langle t_B \rangle^\mc{G}$
being the {\it projection} on a single map 2D phase-space $(x_1,y_1)$
of the {true} {\it four-dimensional} PT function $t_B(x_1,x_2,y_1,y_2)$.
Such estimate, which involves only the shape of our lag-sets once we need to
numerically employ them as recurrence-sets (the dynamics of map (\ref{map2})
at $\beta\neq0$ is fully 4D), is acceptable in the weak-coupling regime, since
the uncoupled case $\beta=0$ has PT function $t_B(x_1,y_1)$ in 2D exactly.
In fact, we expect this approximation to be effective even over very long time-scales,
since the main point here is about the {\it reduction} of departing volume $\mu(D_B)$,
regardless the computational issue of approximating increasingly complex recurrence-sets.\\

\begin{figure}[t!]
  \centering
  \includegraphics[scale=0.88]{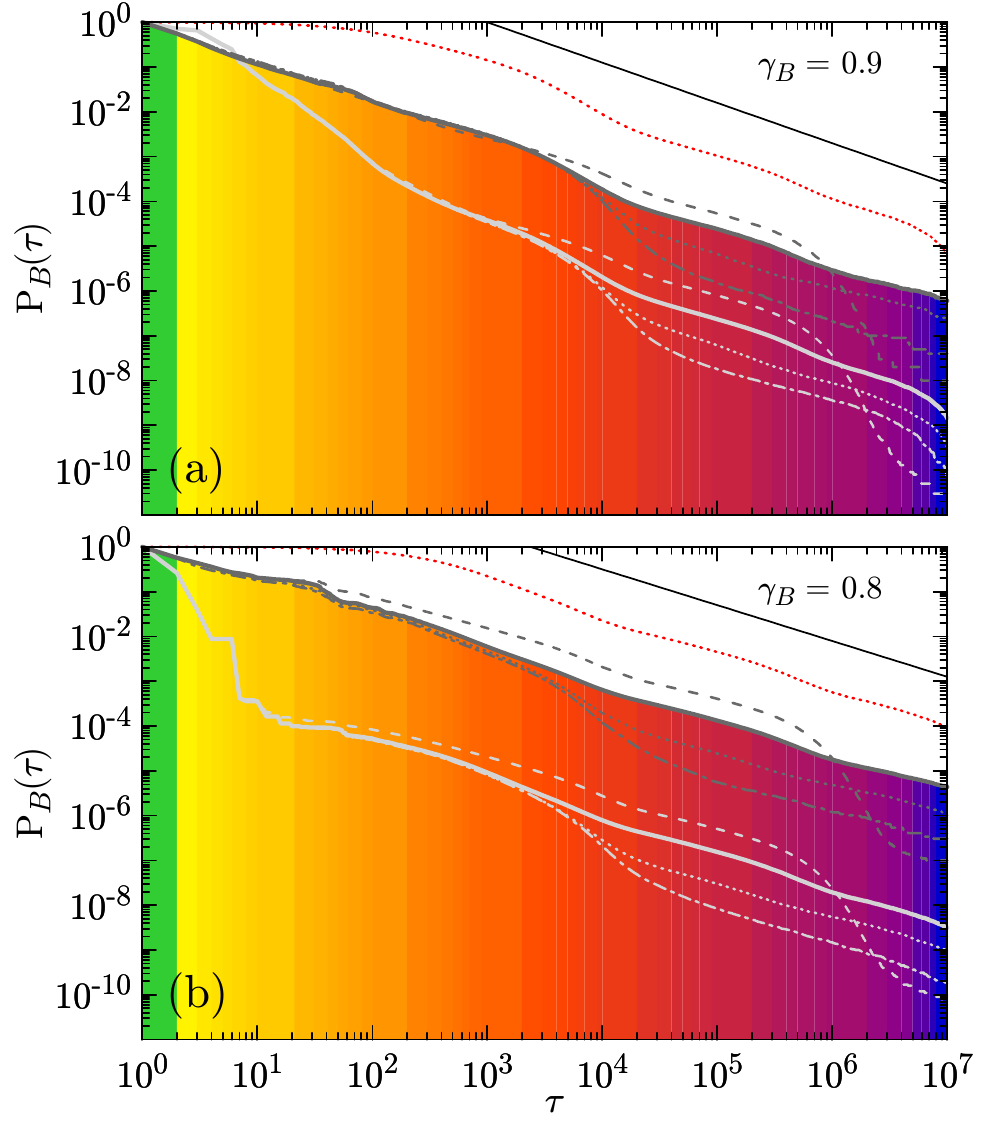}
  \caption{(Color online) RT probability for recurrence-sets
    $B_{[0.1]}$ and $B_{[0.3]}$ (gray curves respectively in panel
    (a), (b)) pictured in Fig.~\ref{fig3} respectively in first and
    third column. The line-types correspond to $\beta=10^{-3}$
    (dashed), $10^{-4}$ (continuous), $10^{-5}$ (dotted) and $10^{-6}$
    (dash-dotted). The two families of gray curves correspond
    to $I^{\delta\tau}(B_{[0.1]})$ (panel (a)) and
    $I^{\delta\tau}(B_{[0.3]})$ (panel (b)) with lag $\delta\tau=200$ 
    and pictured in Fig.~\ref{fig3} respectively in second and fourth
    column. Thin red/dark-gray dotted curves in both panels are
    the light-gray continuous ones ($\beta=10^{-4}$) shifted by
    $\delta\tau$ as in Eq. (\ref{probokI}) for the {\it exact}
    lag-sets. }  
  \label{fig5}
\end{figure}

Such reasoning is supported by the numerical application of the lagging procedure
$B_{[0.1]}\to I^{\delta\tau}(B_{[0.1]})$ for $\delta\tau=200$ (in Fig.~\ref{fig6}, first column of panels is $B_{[0.1]}$,
second column is $I^{200}(B_{[0.1]})$): by drastically {\it reducing} the departure-set to few grid-cells,
in Fig.~\ref{fig5}~(a) we get the corresponding RT probabilities {\it enhanced} by $\sim2$ orders of magnitude,
but also underestimate formula (\ref{probokI}) by a similar factor
(compare black curves for lag-set $I^{200}(B_{[0.1]})$ with the gray ones for the original $B_{[0.1]}$
and the dotted red/dark-gray curve for its theoretical shift by
$\delta\tau=200$ as given by Eq.~(\ref{probokI})). 
In agreement with the previous case at $\beta=0$ and $\delta\tau=10$
in Fig.~\ref{fig3}~(d), our fixed square-grid $\mc{G}$ is again
insufficient in resolving the lag-set details for such a large lag
$\delta\tau=200$, since the filamentation scale is now of order $\sim
e^{-200\,\Lambda}$ with $\Lambda$ the {\it maximal Lyapunov exponent}
of the system.

These observations lead to a first important conclusion: RT probabilities
can be {enhanced} by {reducing} the size of departure-set to some minimal
computable area (in our case,  few grid-cells) even if the chosen recurrence-set
{\it is not} the exact lag-set of some other set. This makes the departing volume
$\mu(D_B)$ a robust and thus useful parameter in the choice of {\it any} recurrence-set.\\

Interestingly, by inspection of second column in Fig.~\ref{fig6} for $I^{\delta\tau}(B_{[0.1]})$,
one can notice that the pair of period-3 island-chains (see the arrows in Fig.~\ref{fig2}) is not
removed by the lagging procedure, because the return time-scale of such resonances is longer
than the chosen lag $\delta\tau=200$.
Indeed, in Fig.~\ref{fig2}~(c) for $\beta=10^{-3}$, such particular island-chains
still show persistent effects on orbit densities at times $\tau\sim10^8$ larger
than the maximal observed RT $=10^7$, denoting very strong stickiness.
In order to test which is the effect of such period-3 structures on the RT statistics, in
Fig.~\ref{fig6} we {\it forcedly exclude} them from the latter by taking a larger rectangular
set $B_{[0.3]}=\{|x_1|<0.3\}$ and its corresponding lag-set $I^{\delta\tau}(B_{[0.3]})$, again
for a lag-time $\delta\tau=200$ (third column is $B_{[0.3]}$, fourth column is $I^{200}(B_{[0.3]})$).
Set $B_{[0.3]}$ is chosen to overlap the targeted period-3 island-chains and {\it transfer}
their dynamical contribution to its {\it residence-times} statistics \cite{Note5}.
Such property is preserved by all lag-sets $I^{\delta\tau}(B_{[0.3]})$ so that, in
Fig.~\ref{fig5}(b) (light-gray curves for $B_{[0.3]}$ and gray curves for
$I^{200}(B_{[0.3]})$) we observe the {\it same} ergodic transition from panel (a),
but now {\it without} the RT contribution of the period-3 structures.
Quite interestingly, the comparison between panels (a) and (b) in Fig.~\ref{fig5}
(both gray and light-gray curves) highlights two different
power-law decays with appreciably different exponents
$\gamma_{[0.1]}\simeq0.9$ (panel (a)) and $\gamma_{[0.3]}\simeq0.8$
(panel (b)) both extended over the whole observation-window
$\tau\leq10^7$. 
While the empirical ergodic threshold $\beta_*\simeq10^{-4}$ and its
characteristic RT range $10^4<\tau<10^6$ are found to be {\it
  independent} on the choice of recurrence-set (as they should),
different decay rates imply that we are still observing the 
{\it non-asymptotic} part of the RT probability distribution.
In other words, even ignoring the time-scale at which the true
asymptotic regime will appear, we can already tell that it must be
larger than $\tau=10^7$. 
To our knowledge, no existing method (not analytic nor numerical) is
capable to provide such lower-bound information in a completely
general, higher-dimensional setting.\\ 

We remark that our Poincar\'e-time (PT) analysis is different from the one
studied in \cite{ac13,a07}, although the two share the same philosophy.
There, several {\it conditional measures} are constructed by selecting only those
Poincar\'e cycles that have RT falling in some specific time-range, which is picked
{\it empirically} from the RT statistics.
Here, instead, we compute the {\it full} orbit density and the local-time
of {\it all} Poincar\'e cycles to get, by PT, an {\it automatic} partition for phase-space
{\it and} for RT distributions as well.

\section{Conclusions}
\label{conc}
In this paper we investigate the dependence of Poincar\'e recurrence-times (RT)
on the choice of recurrence-set.
By deriving a general expression for the return probabilities in terms of
level-sets of a {\it time-function} (PT) that are easily visualized in phase-space,
we are able to {\it localize} all the dynamical structures
responsible for long recurrences by automatically {\it associating} each of them
to a specific bin of the RT statistics.
We exploit PT as a {\it detector} for extremely small phase-space structures
(such as tiny stability-islands) based on their natural recurrence time-scale.
As a second by-product, PT analysis also suggests a rigorous recipe to {\it enhance} the
RT probabilities of a given set by simply {\it deforming} its shape.\\

By taking as a benchmark-model the process of weak Arnol'd diffusion in a four-dimensional
Hamiltonian map, we discover a novel factor ruling the dependence of RT statistics on
the choice of recurrence-set $B$: namely, the {\it volume} of its {\it departure-set}
$D_B=\mh{f}(B)\bsl B$, quantifying the rate per iteration at which the dynamics
of system $\mh{f}$ extracts/re-injects its orbits from/into a set $B$.
Such indicator adds useful information to the standard checks of what are the 
{\it dynamical structures} ({\it e.g.} un/stable periodic orbits) contained in the chosen recurrence-set.
This is practically implemented through numerical experiments by {\it controlling}
the departing volume $\mu(D_B)$, which leads to concrete examples of
{\it enhanced} RT probabilities, and by {\it forcedly excluding} particular resonant
structures from the recurrence-set.
The conjoint analysis of such two operations turns out to be crucial in identifying
a lower-bound for the {\it transient} time-scale of the system: indeed, depending on
which structures are allowed to contribute to the RT statistics, we demonstrate
the existence of extremely long {power-like} decays whose exponent
$\gamma_B$ {\it depends} on the choice of recurrence-set $B$ for the {\it whole} observed
time-window, which can thus be marked as transient.\\

We remark the generality of such phenomenon, which can be encountered in all systems
endowed with a weak-coupling mechanism between chaotic sub-systems. We also point out that
the same considerations can be applied to recurrence properties of {\it ensembles} of
initial-conditions for many, weakly-coupled identical Hamiltonian systems.

\section*{Acknowledgments}
M.S. thanks CAPES (Brazil) for financial support through post-doctoral
fellowship (PNPD), C.M. thanks CNPq, CAPES and FAPESC (all Brazilian
agencies) for financial support. 

\appendix

\section{\label{hs}Numerical histograms}
When dealing with real data, we write the probability
$\tn{P}_B(\tau)=\sum_{t\geq\tau}^{\infty}\,\rho_B(t)$
to {\it do not return} in $t<\tau$ steps as:
\begin{align}
  \tn{P}_B(\tau)\ =\ 1\ -\ \sum_{t=1}^{\tau-1}\,\rho_B(t)\ ,
  \label{Pnum}
\end{align}
since this is a {\it finite} sum and the normalization condition
$\sum_{t=1}^{\infty}\,\rho_B(t)=1$ is ensured by the Poincar\'e recurrence theorem.
Numerically, the exact RT distribution $\rho_B(\tau)$ is then replaced
by the observed {\it number} of recurrences $r_B(\tau)=E\cdot\rho_B(\tau)$
over an ensemble of size $E$, the current {\it total} number of collected cycles.
As $\rho_B(\tau)$ generates the integrated RT probability $\tn{P}_B(\tau)$,
also $r_B(\tau)$ sums up to the observed number $p_B(\tau)$ of events with $\tn{RT}\geq\tau$ :
\begin{align}
p_B(\tau)\ =\ E-\sum_{t=1}^{\tau-1}\,r_B(t)\ \simeq\ E\cdot \tn{P}_B(\tau)\ ,
  \label{Pnum2}
\end{align}
so that both $r_B(\tau)$ and $p_B(\tau)$ are integer-valued, ranging between 0 and $E$.
Once another Poincar\'e cycle is executed and a new $\tn{RT}=\tau'$ is collected,
we raise by 1 the total number of events $E\mapsto E'=E+1$ and apply the following
update rule for the numerical RT density $r_B$: 
\begin{align}
  r_B(\tau) \ \mapsto\  r'_B(\tau) \ =\ \left\{
    \begin{array}{ll}
      r_B(\tau)+1&\textit{, if}\ \ \tau=\tau'\ ,\\ 
      r_B(\tau)&\textit{, else.}
    \end{array}
  \right.
  \label{Nr}
\end{align}
This produces a frequency histograms for observable $\tau$, the RT,
and allows to generate the numeric RT statistics
$p_B(\tau)/E\simeq \tn{P}_B(\tau)$ at {\it finite} number of events $E$,
as a post-process, by Eq.~(\ref{Pnum2}), {\it after} the data collection is over.\\
In the formal limit of an infinite number of events: 
\begin{align}
\lim_{E\to\infty}p_{B}(\tau)/E\ =\ \tn{P}_B(\tau)\ . 
\label{limNp}
\end{align}
one recovers the  RT probability $\tn{P}_B(\tau)=\tn{Prob}(\tn{RT}\geq\tau)$.

\section{Departures \& Arrivals\label{DA}}
All points {\it outside} set $B$ belong to the difference
$\Omega\bsl B$, obtained by removing from phase-space $\Omega$ all points in $B$.
By the volume-preserving character of $\mh{f}$, it follows that phase-space is invariant,
$\mh{f}(\Omega)=\Omega$, so that, by the general property for which:
$\mh{f}(A\bsl B)=\mh{f}(A)\bsl\mh{f}(B)$ for any pair of sets $A$, $B$ (the image of a
difference is the difference of the images), we obtain that all points outside $B$ go,
in one iterate, in the set $\mh{f}(\Omega\bsl B)=\Omega\bsl B_1$, with $B_1\equiv\mh{f}(B)$.
We wish to find the expression for the {\it arrival-set} $A_B$, made of points that,
in a single iteration, arrive in $B$ from its outside; intuitively, this corresponds to
the intersection between the set found above and $B$ itself: $A_B=B\cap(\Omega\bsl B_1)$.
By the general property of set-intersections: $A\cap(B\bsl C)=(A\cap B)\bsl C$
for any triple $A$, $B$, $C$, and the trivial fact that $B\cap\Omega=B$ (since
$\Omega$ contains {\it any} admissible set $B$), we get the desired expression:
\begin{align}
 A_B\ =\ B\cap(\Omega\bsl B_1)\ =\ (B\cap\Omega)\bsl B_1\ =\ B\bsl B_1\ .
\label{arrset}
\end{align}
The basic properties of Lebesgue measure $\mu$, for which $\mu(A\cap B)=\mu(B\cap A)$ and
$\mu(A\bsl B)=\mu(A)-\mu(A\cap B)$ for any pair $A$, $B$, then allow to
express the {\it arriving volume}:
\begin{align}
 \mu(A_B)\ =\ \mu(B)-\mu(B\cap B_1)\ .
\label{arrsetvol}
\end{align}
By repeating the same for the departure-set $D_B=B_1\bsl B$ described in first paragraph of
Sec.~\ref{dc} and using again the volume-preserving property $\mu(B_1)=\mu(B)$ we obtain:
\begin{align}
 \mu(D_B)\ =\ \mu(B_1)-\mu(B_1\cap B)\ =\ \mu(A_B)\ .
\label{depsetvol}
\end{align}
Relation $ \mu(D_B)= \mu(A_B)$ thus tells that the volume of points that
{\it escape} $B$ in one iteration {\it coincides} with the volume that enters in $B$ from its outside.
By the set-properties enlisted above, one can also prove that the arrival- and departure-set
are {\it disjoint} $D_B\cap A_B=\emptyset$.
By similar considerations, one can also find the sets $RES_B=B\cap B_1$, where {\it all}
residence cycles take place, and $REC_B=\Omega\bsl(B\cup B_1)$, where all Poincar\'e
recurrence cycles take place (respectively dark-red/dark-gray
and white zones in Fig.~\ref{fig1}). 
Interestingly, the four sets $D_B$, $REC_B$, $A_B$ and $RES_B$ are all mutually disjoint,
while their union make up the {\it whole} phase-space $\Omega$. They thus form a covering partition of
phase-space which is {\it uniquely} induced by the dynamical
processes of departure, recurrence, arrival and residence for the chosen recurrence-set $B$.

\section{\label{ptf}Poincar\'e-time (PT)}
While counting RT's, we keep track of {\it where} each Poincar\'e cycle
goes by collecting in phase-space the clock function $t_B(\mh{x})$,
named {\it Poincar\'e-time} (PT). This clock is defined as the {\it elapsed-time} since
last exit from $B$ and is localized at orbit position $\mh{x}$. In practice,
we partition a selected two-dimensional sub-space by a square-grid $\mc{G}$
({\it e.g.} in Sec.~\ref{res} we take $\mc{G}$ over {\it one} of the two coupled
2D phase-spaces). By naming $\bs{\chi}_t=\bs{\chi}(\mh{x}_t)\in\mb{N}^2$
the discrete (integer) coordinates of the grid-cell in which point $\mh{x}_t$
falls, $n^\mc{G}(\bschi)$ and $t^\mc{G}_p(\bschi)$ are respectively the 
orbit-density and the PT function {\it summed} over all points falling in cell
$\bschi$. At each step of a cycle, we {\it evolve} the 
orbit point $\mh{x}_{t-1}\mapsto\mh{x}_{t}$ and its discrete
representation $\bs{\chi}_{t-1}\mapsto\bs{\chi}_{t}$ while {\it
  updating} both $\mc{G}$-summed density and PT function: 
\begin{align}
  n^\mc{G}(\bs{\chi}_{t})\ &\mapsto\ n^\mc{G}(\bs{\chi}_{t})+1\ ,\\
  t^\mc{G}_B(\bs{\chi}_t)\ &\mapsto\ t^\mc{G}_B(\bs{\chi}_{t})+t\ .
\end{align}
By collecting ensembles of cycles, we approximate the $\mc{G}$-{\it averaged}
Poincar\'e-time function $\langle t_B\rangle_\mc{G}(\bschi)$ by the ratio:  
\begin{align}
  \langle t_B\rangle_\mc{G}(\bschi)\ =\
  \frac{t^\mc{G}_B(\bschi)}{n^\mc{G}(\bschi)}\ . 
  \label{cellave}
\end{align}
For 2D systems, the $\mc{G}$-averaged PT function $\langle t_B\rangle_\mc{G}(\bschi)$
above coincides with the true PT function $t_B(\mh{x})$, and any error in its
approximation may only come by the representation of smooth boundaries by grid-cells of $\mc{G}$.
Once we consider systems in $N$ dimensions with $N>2$, the true PT function $t_B(\mh{x})$
depends on all $N$ variables, while the $\mc{G}$-average $\langle t_B\rangle_\mc{G}(\bschi)$
still has two variables and corresponds to the projection of the true PT function
further averaged over the $N-2$ dimension {\it not included} in the observed 2D sub-space
where $\langle t_B\rangle_\mc{G}(\bschi)$ is defined.\\
This means that, in 2D, we observe the PT {\it level-sets} $S^\tau_B=\{\ \mh{x}\in \Omega\bsl B\quad\tn{s.t.}\quad t_B(\mh{x})=\tau\ \}$
discretized over grid $\mc{G}$ and thus uniform PT values, constant inside each $S^\tau_B$,
while in higher-dimensions we get continuous values caused by projecting many
different level-sets on grid $\mc{G}$.
To clearly understand the relationship between PT level-sets and RT statistics,
we consider also the update rule, analogous to Eq.~(\ref{Nr}), for
the numeric RT statistics $p_B$:
\begin{align}
  p_B(\tau) \ \mapsto\  p'_B(\tau) \ =\ \left\{
    \begin{array}{ll}
      p_B(\tau)+1&\textit{, if}\ \ \tau\leq\tau'\ ,\\
      p_B(\tau)&\textit{, else.}
    \end{array}
  \right.
  \label{Cr}
\end{align}
for which {\it all} bins from 1 to $\tau'$ in $p_B$ are raised by 1 when an event
$\tau'$ is detected. On the other hand, during a Poincar\'e cycle $\tau$
the PT clock function $t_B(\mh{x})$ takes all values from 1 to $\tau$; by
considering {\it each point} of the cycle as an {\it event} for the numerical
PT {\it probability density} $r_{t_B}(t)$, after a new RT event $\tau'$ is collected,
we get all bins from 1 to $\tau'$ raised by 1. But this is exactly the {\it same}
operation encoded in update rule~(\ref{Cr}) for RT statistics:
\begin{align}
  r_{t_B}(t) \,\mapsto\, r'_{t_B}(t) =\left\{
    \begin{array}{ll}
      r_{t_B}(t)+1&\textit{if}\ \ t\leq\tau'\ ,\\
      r_{t_B}(t)&\textit{else.}
    \end{array}
  \right.
  \label{Np}
\end{align}
Since both histograms $p_B$ and $r_{t_B}$ are set to zero when the experiment starts,
we conclude that $r_{t_B}(\tau)=p_B(\tau)$ for all $\tau$, {\it i.e.} the PT probability {density}
(among {all} steps of all cycles) coincides with the RT probability (among all cycles).
Notably, Eq.~(\ref{Np}) corresponds to a Monte-Carlo approximation of volumes $\mu(S^\tau_B)$ of
the {\it level-sets} $S^\tau_B$ of PT function $t_B(\mh{x})$ which, we can thus already deduce,
are proportional to $\tn{P}_B(\tau)$ up to normalizations.

\section{\label{cp}Cumulative probability}
Our numerical experiments collect branches of orbits that depart
from set $D_B=B_1\bsl B$, {exit} and then arrive in set $A_B=B\bsl B_1$
spending $\tau\geq1$ iterations outside $B$. The RT probability
$\tn{P}_B(\tau)=\tn{Prob}(RT\geq\tau)$ to return in any number
$t\geq\tau$ of iterations coincides with the probability of {\it not}
returning during the first $t<\tau$ steps. Interestingly, the set
of points $S^\tau_B$ that after $\tau$ steps {did not} return to $B$
is obtained by {\it iterating} the previous set $S^{\tau-1}_B$ that did
not return after $\tau-1$ steps and then {\it removing} all points that
arrived in $B$. This leads to the recurrence relation: 
\begin{align}
S^\tau_B\ =\ \mh{f}\left(\,S^{\tau-1}_B\,\right)\ \bsl\ B\ .
\label{recrel}
\end{align}
Since the set of points that do not return in $B$ after $\tau=1$ iterations
is just the departure-set $B_1\bsl B \equiv D_B$, we set  $S^1_B =  D_B$
as the initial condition for relation~(\ref{recrel}) at $\tau=1$. This leads
to the intuitive explicit solution: 
\begin{align}
S^\tau_B\ =\ B_\tau\,\bsl\,\bigcup_{t=0}^{\tau-1}\,B_t\ ,
\label{solset}
\end{align}
which tells that the set of points $S^\tau_B$ not returning to $B$ in the
first $\tau$ steps is just the $\tau$-th iterate of $B$ with all the
previous iterates {\it removed}, including set $B$ itself.
By initial assumption, the volumes of sets~(\ref{solset}) quantify the
RT probability, so that $\mu(S^\tau_B)\propto\tn{P}_B(\tau)$ for all $\tau$;
since one must also have $\tn{P}_B(1)=1$ for normalization, the full RT probability
has the unique expression:
\begin{align}
  \tn{P}_B(\tau)\ =\ \mu\left(\,S^\tau_B\,\right)\ /\
  \mu\left(\,S^1_B\,\right)\ ,  
  \label{probok}
\end{align}
in terms of the set volumes $\mu(S_B^\tau)$. By labelling through the PT function
$t_B(\mh{x}_t)\equiv t$ each of the iterates along each Poincar\'e cycle, we obtain
a phase-space function $t_B(\mh{x})$ whose {\it level-sets}
$S^\tau_B\equiv\{\ \mh{x}\in \Omega\bsl B\quad\tn{s.t.}\quad t_B(\mh{x})=\tau\ \}$
as in Eq.~(\ref{PTlev}) coincide exactly with the sets in Eq.~(\ref{solset}),
since one finds $t_B(\mh{x})=\tau$ in all points in $B_\tau$ that {\it has not been already}
labelled at any previous time $t=0\ldots\tau-1$. This proves that expression (\ref{probok})
for RT statistics $\tn{P}_B(\tau)$ (or, equivalently, Eq.~(\ref{probok1})) corresponds
to the ratio between the $\tau$-th and the first PT level-set volume, as suggested
by comparison of Eq.s~(\ref{Cr}) and (\ref{Np}) from Appendix~\ref{ptf}.



%

\end{document}